\documentclass[aps,pre,groupedaddress,showpacs,twocolumn]{revtex4}
\usepackage{graphicx}

\newcommand{\fsolvas}{f_\mathrm{solv}^{\mathrm{AS}}}
\newcommand{\fsolvs}{f_\mathrm{solv}^{\mathrm{S}}}
\newcommand{\fsolv}{f_\mathrm{solv}}
\newcommand{\fb}{f_\mathrm{b}}
\newcommand{\fint}{f_\mathrm{int}}
\newcommand{\Tast}{T^\ast}
\newcommand{\Tw}{T_\mathrm{w}}
\newcommand{\Tc}{T_\mathrm{c}}
\newcommand{\kb}{k_\mathrm{B}}
\newcommand{\rme}{\mathrm{e}}

\newcommand{\yast}{y^\ast}

\begin{document}
\title{Scaling of solvation force in 2D Ising strip}
\author{Piotr Nowakowski}
\email{pionow@fuw.edu.pl}
\author{Marek \surname{Napi\'orkowski}}
\affiliation{Faculty of Physics, University of Warsaw, ul. Ho\.za 69,
  00-681 Warsaw, Poland} 
\date{\today}
\begin{abstract}
The solvation force for the 2D Ising strip is calculated via
exact diagonalization of the transfer matrix in two cases: the symmetric case corresponds to 
 identical surface fields, and the antisymmetric case to exactly opposite surface fields. 
 In the symmetric case the solvation force is
always negative (attractive) while in the antisymmetric case the solvation force is positive (repulsive) 
at high temperatures and negative at low temperatures. It changes sign close to the 
critical wetting temperature characterizing the semi--infinite system. 
The properties of the solvation force are discussed and the 
scaling function describing its dependence on temperature, surface field, and strip's width is proposed.  
\end{abstract}
\maketitle
\section{Introduction}
Properties of the solvation force in various condensed matter systems have been the subject 
of very intensive research during recent years 
\cite{EvansMarconi, Christ1, Evans1, H1, EvansStecki, HSED, Evans2, KPH, SHD, Nature}. 
Lattice models play special role among considered systems.  Although many important results were 
obtained in this field 
by different methods of taking into account fluctuations which determine the properties of 
the analyzed systems, the 
approach based on the exact evaluation of the partition function via the transfer matrix method still
plays a distinguished role. Below we report our results on the properties of the solvation force for 
2D Ising strip. They are obtained via exact diagonalization of the transfer matrix which is then followed by 
numerical solutions of equations for eigenvalues. 

We consider Ising model on a two--dimensional square lattice consisting of 
$M$ rows and $N$ columns. There is no bulk magnetic field acting on the system 
and there are two surface fields: $h_1$ acts on spins in the first row and $h_2$ acts on the $M$-th 
row. The Hamiltonian of this model takes the standard form 
\begin{eqnarray}
\nonumber\mathcal{H}\mathbf{(}\left\{s_i\right\}\mathbf{)} &=& -J
\sum_{n=1}^{N}\sum_{m=1}^{M-1}s_{n,m} s_{n,m+1}\\ 
\nonumber& &-J\sum_{n=1}^N\sum_{m=1}^M s_{n,m}s_{n+1,m} \\
\label{hamiltonian}& &-h_1\sum_{n=1}^N s_{n,1}-h_2\sum_{n=1}^N s_{n,M},
\end{eqnarray}
where $s_{n,m}=\pm1$ with $n=1,\ldots, N,\  m=1,\ldots,M$ denotes the spin located in the 
$n$-th column and $m$-th row, and $J$ is the coupling constant. Periodic
boundary conditions in the horizontal direction are imposed: 
$s_{N+1,m}\equiv s_{1,m}$. 

Our purpose is to determine the properties of the solvation force experienced by the system boundaries. 
In the following we consider two cases: the symmetric case (S) corresponds to $h_1=h_2$, and
for the antisymmetric case (AS) one has $h_1=-h_2$. 

Although the strip of finite width experiences no phase transition, we shall often refer to two 
critical temperatures: the bulk critical temperature 
$ \kb \Tc = 2J/\ln\left(1+\sqrt{2}\right)$, 
and the wetting temperature 
$\Tw$ which characterizes the critical wetting transition in a semi--infinite system with surface field $h_{1}$. 
The wetting temperature fulfills the equation 
\begin{equation}
\cosh 2K_1=\cosh 2K-\sinh 2K \rme^{-2K},
\end{equation}
where $K=J/\left(\kb\Tw\right)$, $K_1=h_1/\left(\kb\Tw\right)$, and
$\kb$ is the Boltzmann constant. 
For a small surface field $h_1$ this equation leads to 
\begin{equation}
\label{wettemp}\frac{\Tc-\Tw}{\Tc}=\frac{1}{4}\left(1+\sqrt{2}\right)\,\ln\left(1+\sqrt{2} \right)
  \left(\frac{h_1}{J}\right)^2, \quad \frac{h_1}{J} \to 0.
\end{equation}
For $h_1\geq J$ one has $\Tw=0$ and there is no wetting transition. 

\section{Solvation force}
The dimensionless free energy per one column is defined as
\begin{equation}
f\left(T,h_1,M\right)=\lim_{N\to\infty} \frac{1}{N} \, 
\frac{\mathcal{F}}{\kb T} = -\lim_{N\to\infty}\frac{\ln \mathcal{Z}}{N},
\end{equation}
where $\mathcal{Z}$ is the canonical partition function for the
Hamiltonian (\ref{hamiltonian}) which we evaluate using the exact transfer
matrix method \cite{Kaufman,Abraham}. The free energy may be separated into three types of
contributions 
\begin{equation}
f\left(T,h_1,M\right)=M
\fb\left(T\right)+f_\mathrm{s}\left(T,h_1\right)+\fint\left(T,h_1,M\right), 
\end{equation}
where $\fb$ is the bulk free energy per one spin \cite{Onsager} and  
$f_\mathrm{s}$ is the surface free 
energy; both $\fb$ and $f_\mathrm{s}$ are $M$--independent. The remaining term $\fint$ 
describes the interaction between the system boundaries. It tends to 0 as $M$ goes to infinity 
and from this term the solvation force originates. 

The solvation force is, in general, defined as the minus derivative of $\fint$
wrt $M$. In our case, because $M$ takes only integer values, we use the definition 
\begin{equation}
\fsolv\left(T,h_1,M\right)\!=\!-\left[\fint\!\left(T,h_1,M+1\right)-
    \fint\!\left(T,h_1,M\right)\right],
\end{equation}
which leads to the following expression 
\begin{equation}
\fsolv\left(T,h_1,M\right)=f\left(T,h_1,M\right)-f\left(T,h_1,M+1\right)-
  \fb.
\end{equation}

The solvation force for the 2D strip has been already analyzed in the $\Tw=0$ case
\cite{EvansStecki} corresponding to $h_{1}=J$. Our analysis covers the whole spectrum 
$\Tw \ge 0$, i.e. $h_1 \le J$. To calculate the solvation force we used the 
methods described in \cite{Kaufman, Abraham, Stecki, Maciolek}. The complete 
analysis (including also the inhomogeneous boundary fields) will be published elsewhere 
\cite{NN}; here we discuss only the main results. 

\begin{figure}[t]
\includegraphics{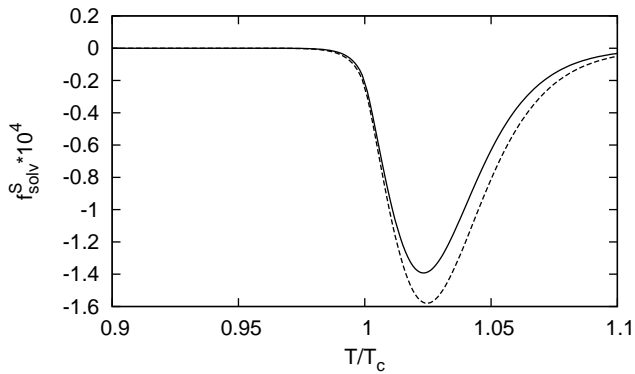}
\caption{\label{fig1} Comparison of the solvation force $\fsolvs$ for $\Tw= 0.8 \Tc$ 
($h_1 \approx
  0.60 J$) (solid curve) and $\Tw=0$ ($h_1=J$) (broken curve) for symmetric case $h_{1}=h_{2}$. 
Both curves correspond to the value $M=50$.}
\end{figure}

First we briefly discuss the symmetric case (S). The corresponding solvation force $\fsolvs$ 
is plotted in Fig.\ref{fig1} for two cases: 
$\Tw > 0$ (solid curve) and $\Tw=0$ (broken curve). The difference
between these two functions is only of quantitative nature --- 
decreasing the surface field $h_{1}$, i.e. increasing the wetting temperature $\Tw$ 
results in decreasing the absolute value of the solvation force. The minimum of 
the solvation force $\fsolvs$ is located at $T_\mathrm{min}^\mathrm{S} > \Tc$. 

\begin{figure}[t]
\includegraphics{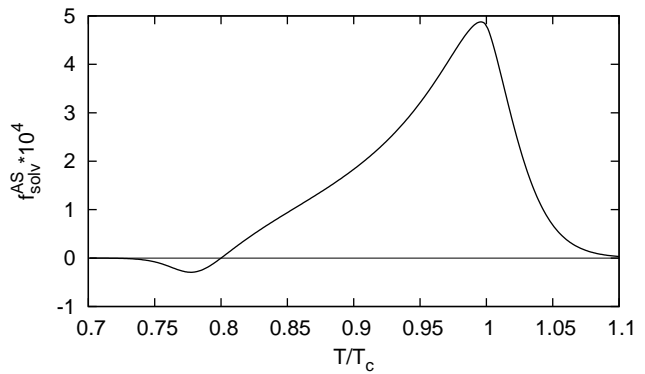}
\caption{\label{fig2}The solvation force $\fsolvas$ as a function of temperature for
  $\Tw=0.8 \Tc$ (i.e. $h_1\approx 0.60 J$) and $M=50$ in the 
  antisymmetric case $h_1=-h_2$. Note that the difference between the zero
  of the solvation force $\Tast$ and the wetting temperature $\Tw$ is not visible in this
  scale.} 
\end{figure}

For opposite surface fields (AS) the solvation force $\fsolvas$ in the $\Tw>0$ case, i.e., $h_1<J$,  
differs substantially from the $\Tw=0$, i.e., $h_1=J$ case. Fig.\ref{fig2} presents a
typical plot of $\fsolvas$ as a function of $T$. For
low temperatures this force is negative (attractive) and has a minimum at $T_\mathrm{min}^\mathrm{AS}<\Tw$. The solvation force is negative at wetting temperature and
has a zero at $\Tast>\Tw.$ Above $\Tast$ the solvation force is positive
(repulsive) and has a maximum at $T_\mathrm{max}^\mathrm{AS}<\Tc$ (for $M$ large enough). This remains 
in contrast with the $\Tw=0$ case in which the solvation force is positive for all
temperatures. 

Exactly at the bulk critical temperature $\Tc$ the dependence of the
solvation force on $M$ has been found using conformal invariance \cite{Nightingale,Cardy} 
\begin{eqnarray}
\fsolvs \left(\Tc,h_1,M\right)=&-\frac{\pi}{48M^2}+\mathcal{O}\left(1/M^3\right),\\
\label{casimir}\fsolvas \left(\Tc,h_1,M\right)=&\frac{23\pi}{48
  M^2}+\mathcal{O}\left(1/M^3\right).
\end{eqnarray}
This result, in particular the universal values of the amplitudes, has been
reproduced by our analysis. 

In the rest of this paper we  exclusively discuss the antisymmetric case (AS). 

First we concentrate on the temperature $\Tast$ at which the solvation
force becomes zero. We have found that for fixed $h_1> 0$ and  for 
$M\to\infty$ this temperature approaches the wetting temperature $\Tw$ exponentially 
\begin{equation}
\label{tast}\frac{\Tast-\Tw}{\Tc}=A\left(h_1\right) \rme^{-B\left(h_1\right)M},\quad M\to\infty,
\end{equation}
where $A\left(h_1\right)$ and $B\left(h_1\right)$ are positive functions of the surface field $h_1$ 
\footnote{In principle, these functions can be found by fitting Eq.(\ref{tast}) to our
numerical results. However, for $\Tw>0.5 \Tc$ and for $M\geq 10$ the
difference between $\Tast$ and $\Tw$ is of order of numerical errors
and we were unable to precisely determine the functions $A\left(h_1\right)$ and $B\left(h_1\right)$.}.
We have found that for $h_1\to 0$, i.e., $\Tw \to \Tc$ one has 
\begin{equation}
\lim_{h_1\to 0} A\left( h_1\right)=0.
\end{equation}

We note that this result is different from the corresponding result obtained within the 
mean field theory where $\Tast$ is exponentially shifted
\textit{below} $\Tw$. It is also different form the corresponding result obtained for the 
restricted
solid--on--solid (RSOS) model, where $\Tast$ is \textit{equal}
to $\Tw$ \cite{ParryEvans}. 

\section{Scaling function}

In this section we discuss the  scaling function that describes the
behavior of the solvation force $\fsolvas(T,h_1,M)$ for large $M$ and subcritical 
temperatures. The relevant scaling function $\mathcal{X}(x)$ has already been
proposed to describe $\fsolvas$ in the $\Tw=0$ case \cite{EvansStecki}
\begin{equation}
\label{fscaleX}\fsolvas\left(T,h_1=J,M\right)=\frac{1}{M^2}
  \mathcal{X}\left(\frac{M}{\xi^-\left(T\right)}\right).
\end{equation}
The correlation length for 2D Ising model close to $\Tc$ is $\xi^-\left(T\right)=\xi_0^- t^{-1}$, where $t=\left(\Tc-T\right)/\Tc>0$ and $\xi^-_0=\left[4
  \ln\left(1+\sqrt{2}\right)\right]^{-1}$. The scaling function $\mathcal{X}(x)$ can
be obtained numerically from the transfer matrix spectrum and some of its properties can
be proved analytically (see Eq.(\ref{casimir}) and \cite{Danchev}), namely 
\begin{eqnarray}
\mathcal{X}\left(0\right)&=&23\pi/48,\\
\mathcal{X}\left(x\right)&=&2\pi^2/x, \quad \text{for }x\to\infty.
\end{eqnarray}
Here we would like to extend this result to $h_1 < J$, i.e. $\Tw>0$ case.

For 2D Ising model the gap exponent $\Delta_1=1/2$ and  for $T<\Tc$ the following 
scaling behavior 
\begin{equation}
\label{fscale}\fsolvas\left(T,h_1,M\right)=\frac{1}{M^2} \mathcal{Y}\left(x,y\right),
\end{equation}
where 
\begin{equation}
\label{A0}x=\frac{M}{\xi^-_0 t^{-1}},\quad y=\frac{A_0}{\kb\Tc} \frac{h_1}{t^{1/2}}
\end{equation}
comes into play in the limit $M\to \infty$ with $x$ and $y$ fixed. This implies additionally the 
$t \to 0$, $h_1 \to 0$ and $\Tw \to \Tc$ limits.
The coefficient $A_0=\left[\left(1+\sqrt{2}\right)/\ln\left(1+\sqrt{2}\right)\right]^{1/2}$ in Eq.(\ref{A0})
has been introduced such that the value $y=1$ corresponds to $T=\Tw$ and then 
Eq.(\ref{wettemp}) is satisfied. For $y < 1$ Eq.(\ref{fscale}) gives the solvation force $\fsolvas$ 
below the wetting temperature and for $y>1$ above $\Tw$. 

\begin{figure}[t]
\includegraphics{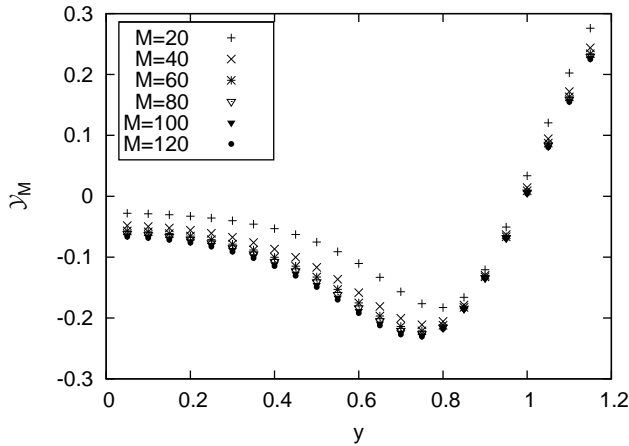}
\caption{\label{fig3}Convergence of the scaling function. The function 
$\mathcal{Y}_M$ is evaluated for $x=1$ and for several values of $y$ and $M$ presented on the plot.}
\end{figure}
\begin{figure}[t]
\includegraphics{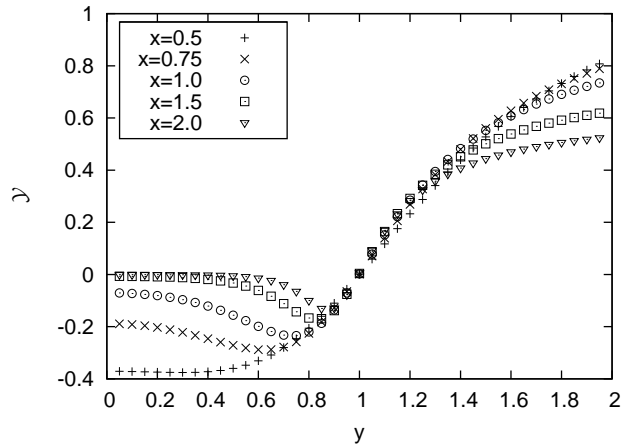}
\caption{\label{fig4}Plots of the scaling function $\mathcal{Y}$. All plots were obtained from 
$\mathcal{Y}_M$ for $M=200$.} 
\end{figure}
\begin{figure}[t]
\includegraphics{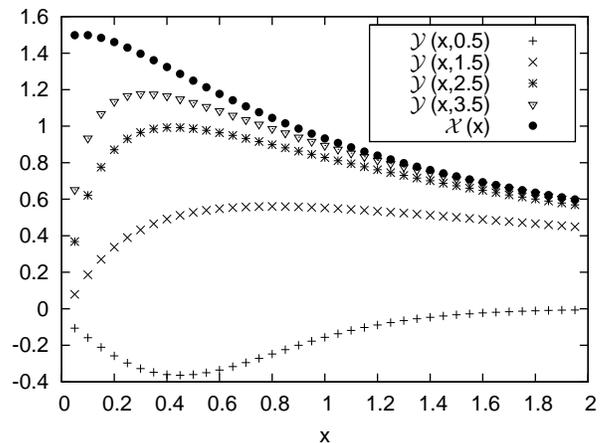}
\caption{\label{fig5}The scaling function $\mathcal{Y}$ plotted for $y=0.5, 1.5,
  2.5, 3.5$ and the scaling function $\mathcal{X}$. All points were calculated
  for $M=200$.}  
\end{figure}

Eq.(\ref{fscale}) may be rewritten in the form leading to the scaling
function
\begin{equation}
\label{limit}\mathcal{Y}\left(x,y\right)=\lim_{M\to\infty}
  \mathcal{Y}_M\left(x,y\right),
\end{equation}
where 
\begin{eqnarray}
\lefteqn{\mathcal{Y}_M\left(x,y\right) = }\\
\nonumber & M^2 \fsolv\left(\Tc\left(1-\frac{x
        \xi_0^-}{M}\right), \kb \Tc y \left(\frac{M}{x \xi_0^-
        A_0^2}\right)^{1/2} , M\right).
\end{eqnarray}
        
Fig.\ref{fig3} presents plots of $\mathcal{Y}_M$ for fixed value of $x=1$ and selected values of 
$y \in [0, 1.2]$ to exhibit the convergence of the series $\mathcal{Y}_M$. Typically, 
for $M$ large enough one has 
\begin{eqnarray}
\label{yyy}
\mathcal{Y}_M\left(
    x,y\right)=\mathcal{Y}\left(x,y\right)+\frac{C\left(x,y\right)}{M}+O\left(M^{-2}\right).
\end{eqnarray}
To estimate the values of the function $\mathcal{Y}$ we evaluated  $\mathcal{Y}_M$ for
$M=200$. We note that a different way of obtaining the function $\mathcal{Y}$, based on the least 
squares method which allows to calculate the functions $\mathcal{Y}$ and $C$ in Eq.(\ref{yyy}), 
leads to similar results (the differences are not visible on the scale
of our figures). The plots of function $\mathcal{Y}$ are shown on Fig.\ref{fig4}. 

Next we investigate the relation between the scaling functions 
$\mathcal{Y}\left(x,y\right)$ and $\mathcal{X}\left(x\right)$. The function 
$\mathcal{X}\left(x\right)$ is calculated in the
limit $M\to\infty$, $t\to 0$ with $h_1$ and $M t$ fixed. By applying
this limit to Eqs (\ref{fscale}) and (\ref{fscaleX}) one~gets 
\begin{eqnarray}
\label{y1y1}
\mathcal{X}\left(x\right)=\lim_{y\to\infty}\mathcal{Y}\left(x,y\right).
\end{eqnarray}
The functions $\mathcal{X}\left(x\right)$ and $\mathcal{Y}\left(x,y\right)$ plotted 
for selected values of $y$ are presented on Fig.\ref{fig5}.
Additionally we have found that $\mathcal{X}\left(x\right)-\mathcal{Y}\left(x,y\right)\propto y^{-2}$ 
for large $y$.

The scaling function $\mathcal{Y}\left(x,y\right)$ changes its sign, see Figs \ref{fig4}, \ref{fig5}. 
The zeros of the scaling function are denoted by $\yast\left(x\right)$, i.e.,  
$\mathcal{Y}\left(x,\yast(x)\right)=0$. 
We have found that for large $x$ the function
$\yast\left(x\right)$ approaches $1$ exponentially form above. This
allows us to show that in the scaling limit $M\to\infty$, $h_1\to 0$, and
$M h_1^2$ fixed one has
\begin{eqnarray}
\nonumber\frac{\Tast-\Tw}{\Tc}&=&\frac{\Tc-\Tw}{\Tc}-\frac{1}{M}
f\left(\frac{M h_1^2}{\kb^2
    \Tc^2}\right)+\mathcal{O}\left(1/M^2\right),\\
\label{func_f}&=& \frac{1}{M} g\left(\frac{M h_1^2}{\kb^2
    \Tc^2}\right)+\mathcal{O}\left(1/M^2\right),
\end{eqnarray}
where $f\left(\zeta\right)$ and $g\left(\zeta\right)$ are positive functions which can be determined via an implicit formula
\begin{eqnarray}
A_0 \zeta^{1/2}&=&\left[f\left(\zeta\right)\right]^{1/2} \yast\left(
  f\left(\zeta\right)/\xi_0^- \right),\\
g\left(\zeta\right) &=& A_0^2-f\left(\zeta\right).
\end{eqnarray}
We note that Eq.(\ref{func_f}) is different form Eq.(\ref{tast}) because different limiting 
procedures were applied in these two cases.
\section{Conclusions}

In the symmetric case of identical surface fields the solvation force is negative (attractive) and 
has minimum at supercritical temperature. The solvation force  
calculated for system parameters such that $\Tw>0$ differs only quantitatively from the one in case 
$\Tw=0$, see Fig.\ref{fig1}.  

In the antisymmetric case of opposite surface fields the solvation
force is positive at high temperatures and negative at low temperatures; 
it changes its sign at temperature $T^\ast > \Tw$, see Fig.\ref{fig2}. In the case $h_1$
fixed and $M \to \infty$ the difference $\Tast-\Tw$  approaches $0$ 
exponentially quickly in $M$. 

The scaling function $\mathcal{Y}(x,y)$ was proposed to describe the behavior of the 
solvation force for $T<\Tc$ in the limit $h_1\to 0$ and $M\to
\infty$, see Figs \ref{fig4}, \ref{fig5}.  We checked that in the limit of high
surface field ($y\to\infty$ in Eq.(\ref{fscale})) this scaling function
approached the scaling function describing the scaling behavior of the for solvation force 
in the $\Tw=0$ case. In addition, the zeros of the scaling function were investigated to find
the formula for $\Tast-\Tw$ in the $h_1\to 0$ limit.

We thank A. Macio{\l}ek, D. Danchev, and S. Dietrich for many helpful discussions. This 
work has been financed from the funds provided by Polish Ministry of Science and Higher Education 
for scientific research for the years 2006--2008 under the research project N202 076 31/0108.

\end{document}